\documentclass[a4paper,preprintnumbers,floatfix,twocolumn,aps,prb,unsortedaddress,superscriptaddress]{revtex4-2}

\usepackage[para]{threeparttable} 
\usepackage{amsmath}
\usepackage{graphicx} 
\usepackage{booktabs}
\usepackage{subcaption}
\usepackage{url}
\usepackage{hyperref}
\usepackage{miller}
\usepackage{bm}
\usepackage{enumitem}
\usepackage{color}

\usepackage[paperwidth=210mm,paperheight=297mm,centering,hmargin=1.7cm,vmargin=2.5cm]{geometry}

\usepackage[utf8]{inputenc}

\begin{document}

\title{Threshold displacement energies in refractory high-entropy alloys}

\author{J. Byggmästar}
\thanks{Corresponding author}
\email{jesper.byggmastar@helsinki.fi}
\affiliation{Department of Physics, P.O. Box 43, FI-00014 University of Helsinki, Finland}
\author{F. Djurabekova}
\affiliation{Department of Physics, P.O. Box 43, FI-00014 University of Helsinki, Finland}
\affiliation{Helsinki Institute of Physics, Helsinki, Finland}
\author{K. Nordlund}
\affiliation{Department of Physics, P.O. Box 43, FI-00014 University of Helsinki, Finland}

\date{\today}

\begin{abstract}
{Refractory high-entropy alloys show promising resistance to irradiation, yet little is known about the fundamental nature of radiation-induced defect formation. Here, we simulate threshold displacement energies in equiatomic MoNbTaVW using an accurate machine-learned interatomic potential, covering the full angular space of crystal directions. The effects of local chemical ordering is assessed by comparing results in randomly ordered and short-range-ordered MoNbTaVW. The average threshold displacement energy in the random alloy is $44.3 \pm 0.15$ eV and slightly higher, $48.6 \pm 0.15$ eV, in the short-range-ordered alloy. Both are significantly lower than in any of the constituent pure metals. We identify the mechanisms of defect creation and find that they are mainly dependent on the masses of the recoiling and colliding elements. Low thresholds are generally found when heavy atoms (W, Ta) displace and replace the lightest atoms (V). The average threshold energies when separated by recoiling element are consequently ordered inversely according to their mass, opposite to the trend in the pure metals where W has by far the highest thresholds. However, the trend in the alloy is reversed when considering the cross sections for defect formation in electron irradiation, due to the mass-dependent recoil energies from the electrons.}
\end{abstract}

\maketitle

\section{Introduction}
\label{sec:intro}

The threshold displacement energy (TDE) is the most fundamental property of radiation damage, defined as the recoiling kinetic energy required to displace an atom to create one or more stable defects (Frenkel pairs)~\cite{nordlund_molecular_2006}. It is the key material-specific parameter needed for models that predict the amount of defects formed in primary radiation damage~\cite{norgett_proposed_1975,nordlund_improving_2018,yang_full_2021}. In an elemental material, the TDE depends only (but strongly) on the crystal direction of the recoiling atom, which in turn also means a dependence on crystal structure. For example, in body-centred cubic metals like W or Fe, both experiments~\cite{maury_frenkel_1978,maury_anisotropy_1976,lomer_anisotropy_1967} and simulations~\cite{nordlund_molecular_2006,byggmastar_machine-learning_2019,bjorkas_modelling_2009,olsson_ab_2016} show that the TDE is lowest for recoils close to the \hkl<100> and \hkl<111> directions, which correspond to head-on collisions with the second- and first-nearest neighbours, respectively. The directional anisotropy is strong, and other crystal directions can have TDEs as much as five times higher than the minimum TDE~\cite{byggmastar_machine-learning_2019}. For primary damage models the most relevant quantity is the TDE averaged uniformly over all crystal directions. This quantity is not directly accessible from experiments and hence require experiment-based approximate models~\cite{jan_zur_1963,jung_anisotropy_1972,maury_frenkel_1978} or molecular dynamics (MD) simulations, either classical~\cite{nordlund_molecular_2006} or \textit{ab initio} MD ~\cite{olsson_ab_2016}, although the latter is computationally expensive.

In an alloy, apart from the directional dependence of the TDE, there is the additional and far-from-trivial dependence on the element of the recoiling atom and its chemical surrounding. This dependence becomes particularly difficult to predict in randomly ordered or many-element alloys such as high-entropy alloys. Even if the TDEs in the pure metals that constitute a given alloy are known, it is not self-evident whether the TDEs in the alloy are higher, lower, or in-between the elemental values. Previous studies on TDEs in alloys are scarce~\cite{gao_point-defect_1993,zhao_alloying_2020,juslin_simulation_2007,ye_primary_2021,wei_effects_2023}, and for high-entropy alloys essentially non-existent~\cite{do_origin_2018}. Gao and Bacon simulated TDEs in Ni$_3$Al and observed a strong dependence on recoiling element~\cite{gao_point-defect_1993}. More recently, Zhao also found strong effects of chemical surrounding in Ni-based binary concentrated solid solution alloys~\cite{zhao_alloying_2020}. In bcc alloys, simulations have shown that the average TDE in steel-like FeCr alloys is similar to pure Fe~\cite{juslin_simulation_2007} and somewhat lower in FeCrAl~\cite{ye_simulation_2023}. Wei et al. simulated the binary W-Mo and W-Ta systems and found that the mass difference (W-Mo) has a stronger effect on the average TDE than atom size mismatch and lattice distortion (W-Ta)~\cite{wei_effects_2023}. Additionally, the importance of using interatomic potentials that are accurately fitted in the range of intermediate to short interatomic distances has been emphasised in several studies~\cite{stoller_impact_2016,byggmastar_effects_2018,bjorkas_modelling_2009,byggmastar_machine-learning_2019}.

Recently~\cite{byggmastar_simple_2022}, we developed a machine-learned potential (tabGAP) for MoNbTaVW and used it to study the behaviour of defects and segregation. A key feature of this potential is that it is accurately fitted to the high-energy repulsion between all elements, which makes it applicable to simulations of radiation damage~\cite{koskenniemi_efficient_2023,wei_effects_2023}. Tungsten-based refractory high-entropy alloys have experimentally shown to be highly radiation-resistant~\cite{el-atwani_outstanding_2019,zong_study_2022,el-atwani_comparison_2023,cheng_enhanced_2023}, yet little is known about the fundamental mechanisms of radiation damage such as the TDE and its dependence on chemical surrounding and crystal direction. It is also computationally well established that MoNbTaVW has a strong preference for short-range ordering~\cite{fernandez-caballero_short-range_2017,byggmastar_modeling_2021}. This adds yet another complication when it comes to understanding the fundamental material properties, including the TDE. The aim of this work is to thoroughly investigate and understand the displacement mechanisms in equiatomic MoNbTaVW, considering all effects that influence the TDE: crystal direction, recoiling element, and chemical surrounding including possible short-range order. 

\section{Methods}
\label{sec:methods}

All molecular dynamics simulations are run with \textsc{lammps}~\cite{thompson_lammps_2022} and the tabGAP Mo--Nb--Ta--V--W potential~\cite{byggmastar_simple_2022}. The threshold displacement energy for a given recoil atom in a given direction is determined by simulating increasingly higher kinetic energy recoils until a stable Frenkel pair is formed. Starting from 10 eV, the recoil energy is increased in steps of 2 eV. Every 500 time steps during the simulation, we use the common neighbour analysis (CNA) algorithm as implemented in \textsc{lammps} to identify whether there are any defects. If there are none (i.e. the CNA identifies all atoms as body-centred cubic), the recoil energy was not enough to displace any atoms, so the simulation is stopped and the next simulation with a higher recoil energy is started. Tests showed that the 500-step ($\lesssim 1.5$ ps, using an adaptive time step with maximum 3 fs) interval was long enough for the displaced atom to leave its lattice site and complete the collision with the neighboring atoms, so that the simulation can safely be stopped if there are no defects as no further defects can possibly be created. If defects are created, the simulation continues until the defects possibly annihilate by returning to their initial positions or relax to new lattice positions during the maximum time of 6 ps. 6 ps is enough to cool down the system sufficiently close to the initial temperature.

Each recoil direction is chosen as a uniformly distributed random point on the unit sphere. The recoil atom is also chosen randomly within a sphere of diameter 7 Å in the centre of the simulation system. To additionally avoid the bias of simulating recoils in the same local chemical environment, the simulation cell is randomly shifted in all three dimensions before every new direction is simulated. To further remove any possible bias, the TDEs are simulated in 10 different simulation boxes with different random order of the atoms. All simulation boxes are relaxed to zero pressure and an initial temperature of 40 K, chosen as a reasonable ``low temperature'' to provide some thermal displacements. A Nosé-Hoover thermostat~\cite{nose_molecular_1984,hoover_canonical_1985} is applied to a 2.5 Å thick border shell while all other atoms evolve in the $NVE$ ensemble. The size of the simulation box is $12 \times 13 \times 14$ unit cells (4368 atoms), which proved to be enough for recoil sequences not to reach the thermally controlled periodic borders. The atoms in the lattice are randomly ordered with the equiatomic MoNbTaVW composition. For comparison, we also repeated the simulation in each of the pure metals. These required slightly larger boxes ($14 \times 15 \times 16$ unit cells, 6720 atoms) due to longer-range rapid recoil sequences especially in the close-packed \hkl<111> directions. In the alloy, the lattice distortion and mass mismatch efficiently limit the range of these \hkl<111> collision sequences. The noncubic simulation cells also ensure that even if some $\hkl<111>$ recoil chains cross the periodic borders, they will not pass through the initial recoil position and cause an artificial Frenkel pair recombination.

Several previous studies have established that at low temperatures some degree of chemical order is favoured in MoNbTaVW and similar alloys~\cite{fernandez-caballero_short-range_2017,yin_ab_2020}. The ordering at low temperatures is strong and is captured accurately by the tabGAP~\cite{byggmastar_modeling_2021}. Hence, in addition to simulating randomly ordered MoNbTaVW, we prepare 10 short-range-ordered MoNbTaVW systems (SRO-HEA) to assess the influence of chemical order on the TDEs. The SRO-HEA systems are prepared by running hybrid Monte Carlo (MC) + MD simulations as implemented in \textsc{lammps}, starting from randomly ordered systems of the same box size as bove (4368 atoms). 10 MC swap attempts for every element pair is performed every 10 MD steps for a total of $4 \times 10^6$ MD steps in the $NPT$ Nosé-Hoover ensemble and $4 \times 10^7$ MC swap attempts. Despite simulating the TDEs at 40 K, when preparing the SRO-HEA systems we set both the MC and the MD temperature to 300 K to create a more experimentally relevant short-range-ordering. The systems are then cooled down to 40 K before the TDE simulations.

\section{Results and discussion}
\label{sec:results}

\subsection{Repulsive potential validation}

\begin{figure}
    \centering
    \includegraphics[width=\linewidth]{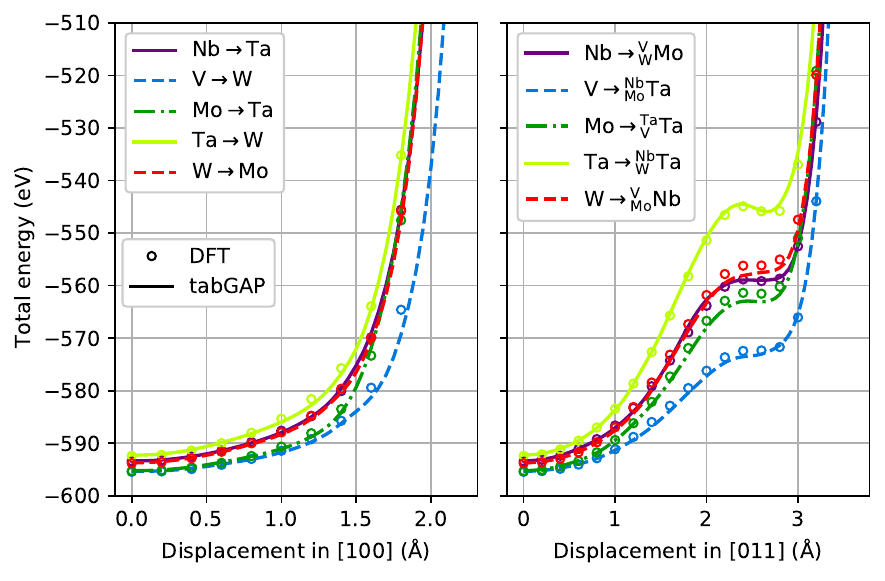}
    \caption{Tests for the accuracy of tabGAP for interatomic repulsion in the HEA. The total potential energy during step-wise movement of one atom in a rigid randomly ordered MoNbTaVW lattice compared with DFT. The legends indicate the element of the displaced atom and the neighbouring atoms it approaches (for the \hkl[011] direction, the atom first passes between two nearest-neighbour atoms as indicated).}
    \label{fig:mbr}
\end{figure}

Before running the TDE simulations, we validated that the tabGAP reproduces interatomic repulsion in the crystal accurately. Fig.~\ref{fig:mbr} shows a direct comparison between density functional theory (DFT) and the tabGAP for the energy landscape when stepwise dragging an atom (of given element) along a \hkl[100] (Fig.~\ref{fig:mbr}a) and \hkl[011] (Fig.~\ref{fig:mbr}b) direction in different randomly ordered rigid lattices. The DFT calculations are done with \textsc{vasp}~\cite{kresse_ab_1993,kresse_efficient_1996} with input consistent with the training data for the tabGAP to allow direct comparison of total energies~\cite{byggmastar_modeling_2021}. The tabGAP follows the DFT curves closely for all the sampled paths. Since these roughly represent the energy landscapes for recoiling atoms in the high-entropy alloy and because the tabGAP is also fitted to vacancy and self-interstitial atom defects~\cite{byggmastar_simple_2022,byggmastar_modeling_2021}, we are confident that the tabGAP can accurately model threshold displacement energy events.

\subsection{Threshold displacement energies}

\begin{figure}
    \centering
    \includegraphics[width=\linewidth]{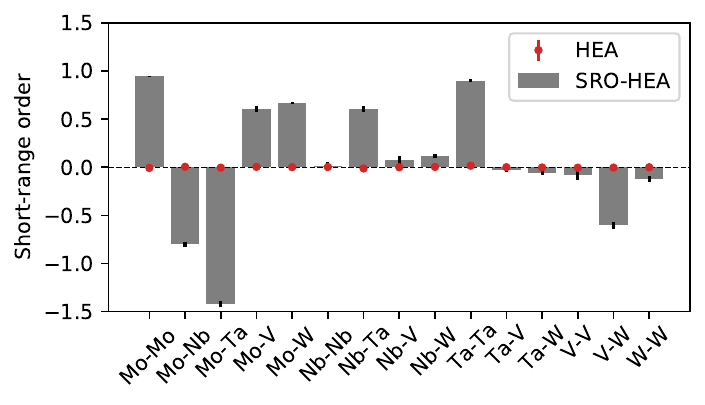}
    \includegraphics[width=\linewidth]{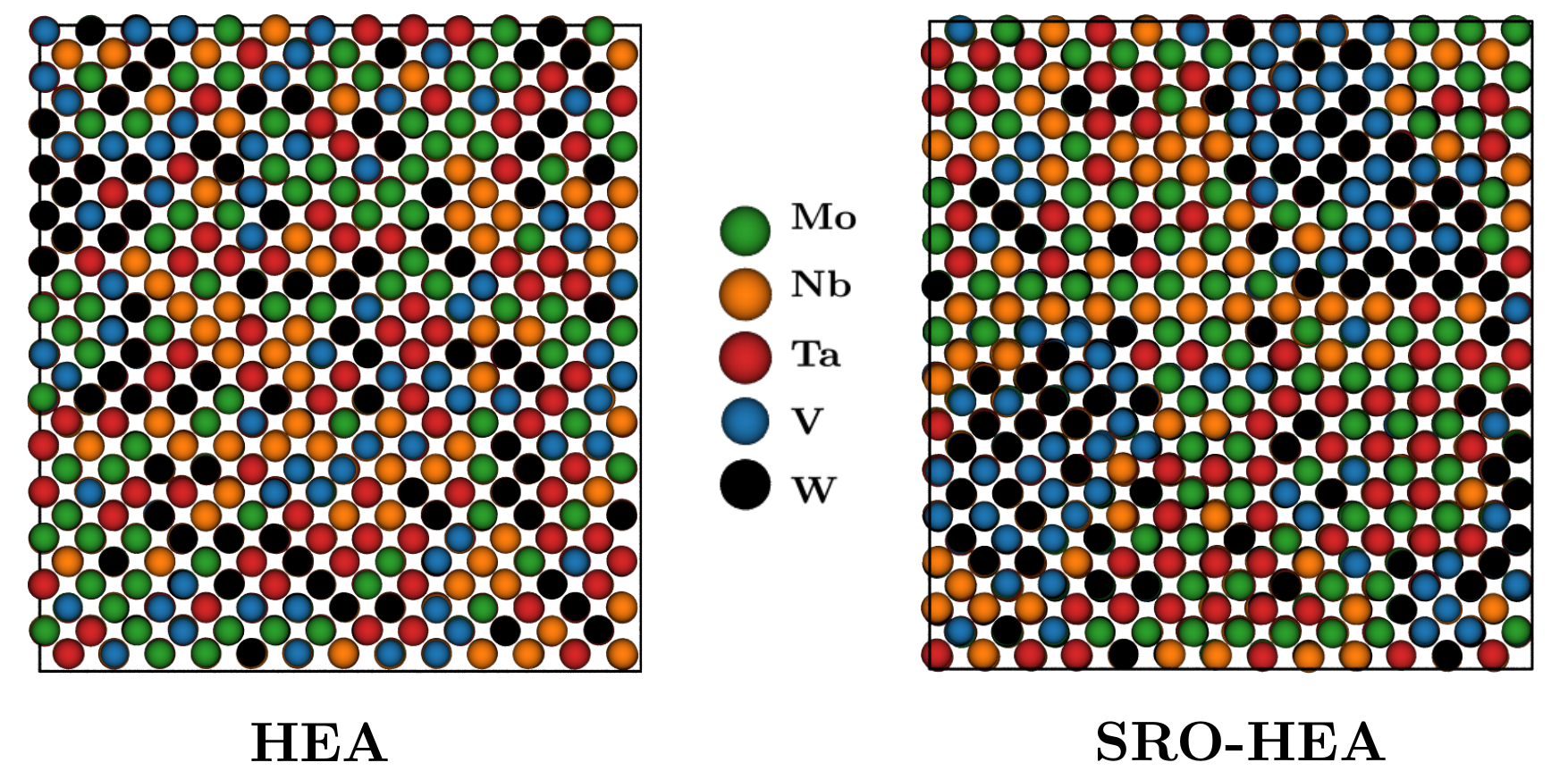}
    \caption{Top: First-nearest-neighbour short-range order parameters in the SRO-HEA and HEA. The values are averages of the ten different simulation boxes, with the standard deviations as error bars. Bottom: \hkl(100) views of one each of the HEA and SRO-HEA boxes.}
    \label{fig:sro}
\end{figure}

Snapshots of the alloy systems and the average short-range order parameters for first-nearest-neighbour pairs in the 10 systems are computed and presented in Fig.~\ref{fig:sro}. As expected, the average SRO parameters for all element pairs in the random HEA are zero, showing that the system completely lacks short-range order. In the SRO-HEA systems, the chemical ordering mainly consists of formation of Mo--Ta, Mo--Nb, and V--W binaries, i.e. local regions of MoTa (mixed with MoNb) and VW ordered binaries just like in our previous work~\cite{byggmastar_modeling_2021}.

\begin{figure}
    \centering
    \includegraphics[width=\linewidth]{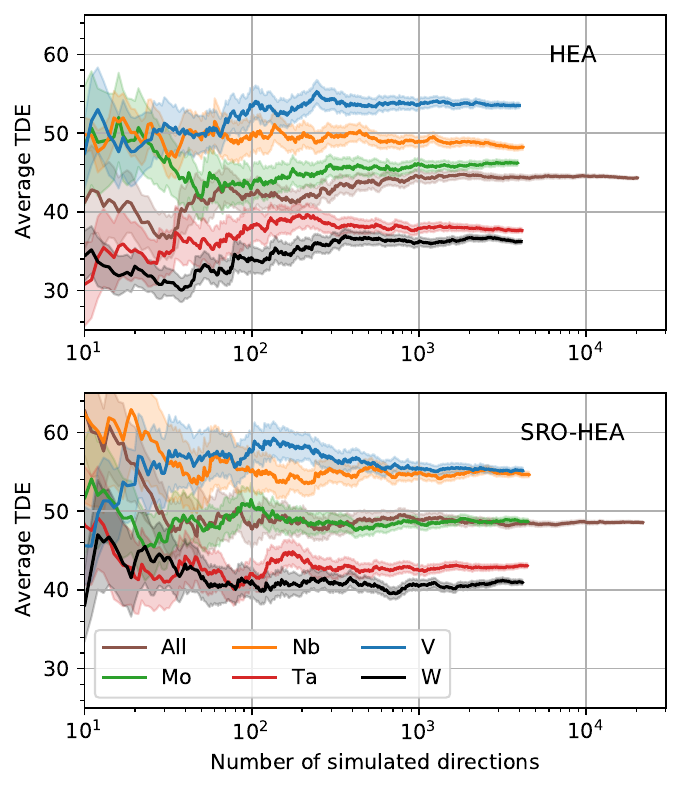}
    \caption{Convergence of the average threshold displacement energy in HEA and SRO-HEA lattices, shown as the cumulative moving averages as functions of number of simulated directions. The data are also separated by recoiling element. The shaded areas show the standard errors of the means.}
    \label{fig:cmlt}
\end{figure}

We ensured that we sample a sufficient number of crystal directions to get converged average TDEs by computing and tracking the cumulative average TDE for each recoiling element as a function of number of simulated directions. Fig.~\ref{fig:cmlt} shows the convergence of the average TDE. After around 500 random directions, the standard error of the mean drops below 1 eV, at around 2000 directions the error is 0.5 eV. Note that the uncertainty of each TDE value is $\pm 1$ eV, since we increase the trial recoil energy in steps of 2 eV. Based on the convergence in Fig.~\ref{fig:cmlt}, we simulated around 4000 directions per recoiling element (in total 20000 simulations per HEA material).

\begin{figure}
    \centering
    \includegraphics[width=\linewidth]{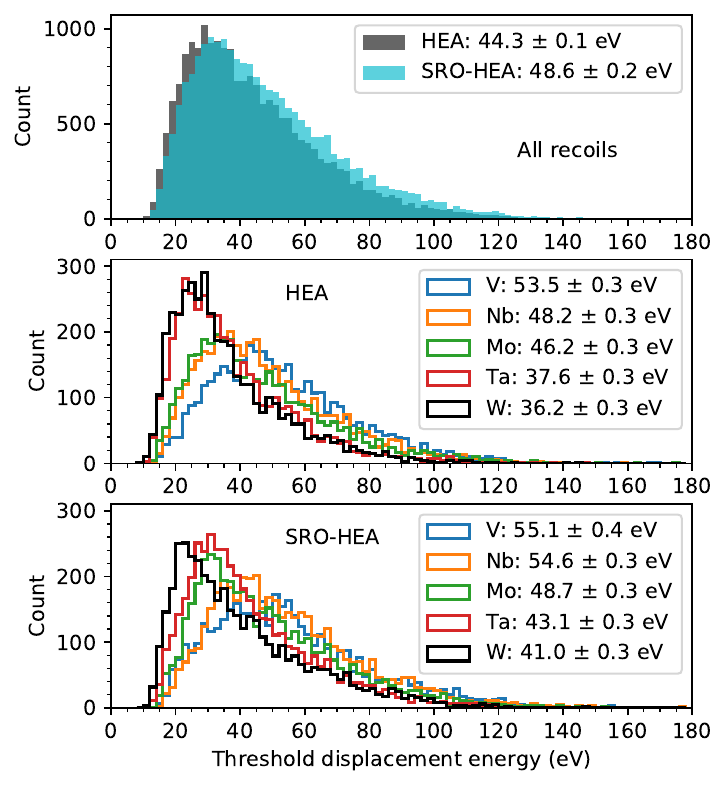}
    \caption{Histograms showing the distributions of threshold displacement energies in the HEA and SRO-HEA. The top figure shows all simulated recoil directions combined. In the lower figures they are separated by recoiling element. The legends show the averages and standard errors of the means.}
    \label{fig:hist}
\end{figure}

Fig.~\ref{fig:hist} shows the distributions of TDEs in the HEA and SRO-HEA, both the total distributions (top) and separated by recoiling element (bottom). The global average TDEs over all directions and elements are indicated in the legend (top), and the averages for each recoiling element (bottom). The distributions are wide, with the minimum TDE as low as 9 eV (although only two data points out of 20293) in the HEA and 11 eV in the SRO-HEA (4 points out of 21828). The effect of short-range ordering is moderate and only produces a small, around 4 eV, shift of the distribution and averages towards higher TDEs. Some additional differences can be observed in the distributions separated by element, where e.g. the distributions of Mo and Ta recoils are shifted towards each other in the SRO-HEA compared to the HEA distributions. This is a consequence of the favourable Mo--Ta nearest-neighbour ordering in the SRO-HEA.

\begin{figure}
    \centering
    \includegraphics[width=\linewidth]{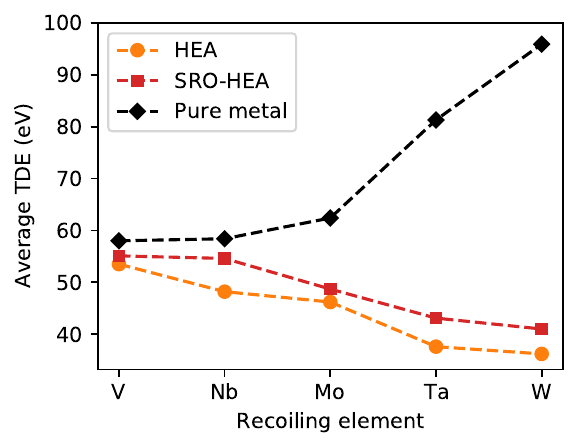}
    \caption{Average threshold displacement energies per recoiling element simulated in the HEA and SRO-HEA, compared to the constituent pure metals. The standard errors of the mean are smaller than the data markers and not visible.}
    \label{fig:avg}
\end{figure}

Fig.~\ref{fig:hist} also shows that the average TDEs separated by elements are ordered inversely with the mass, with V as the lightest atom having the highest average TDE and W the lowest average TDE. This is visualised more clearly in Fig.~\ref{fig:avg}, where we also show the average TDE in each of the pure metals for comparison. Pure W has by far the highest average TDE ($95.9 \pm 1.0$ eV), and the difference to W recoils in the HEA is more than a factor of two. For the lighter elements, the differences are not as dramatic, although it is worth noting that even the element with the highest TDE in the alloys (V) is lower than the lowest pure-element average TDE (also V).

\subsection{Angular anisotropy}

\begin{figure*}
    \centering
    \includegraphics[width=\linewidth]{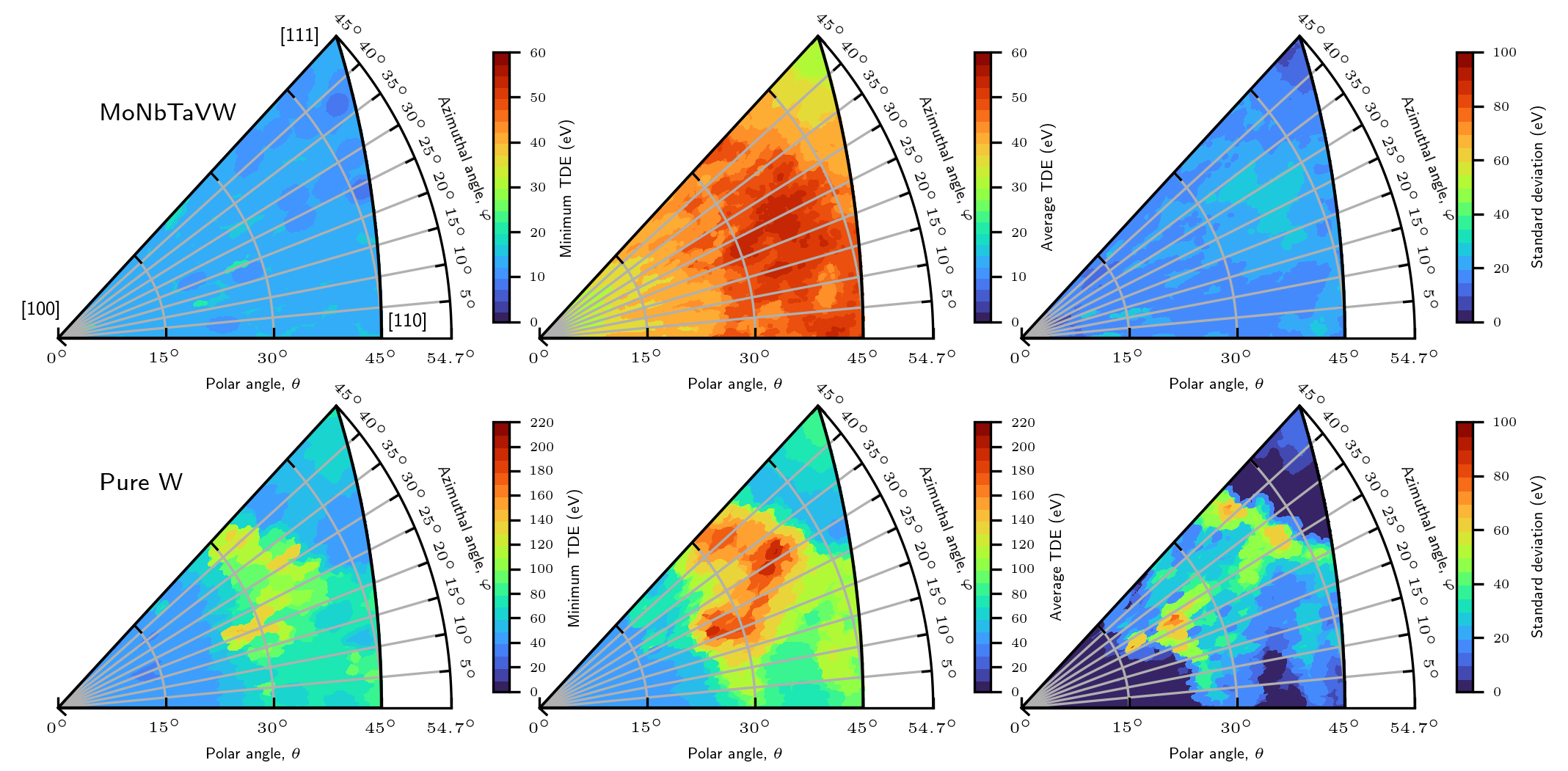}
    \caption{Angular maps of the threshold displacement energies in the HEA (MoNbTaVW, top row) compared with pure W (bottom row), illustrating the weakened anisotropy of the TDE surface in the HEA. The leftmost figures show the minimum TDEs in $2^\circ$-radius regions, the middle figures show the corresponding averages, and the rightmost figures show the standard deviations. Note the different scales of the colour maps.}
    \label{fig:maps}
\end{figure*}

We visualise the angular dependence of the TDE in Fig.~\ref{fig:maps}. Only the random HEA is shown, as the results for SRO-HEA are qualitatively identical. Pure W is shown for comparison. Fig.~\ref{fig:maps} shows both the minimum TDEs, the average TDEs, and the standard deviation as functions of the spherical coordinates. The standard deviation is useful to show the effects of randomness, which in pure W is only thermal displacements, but in the HEA also variations in the chemical environment. The data are reduced by the symmetry of the lattice to the smallest stereographic triangle. In the HEA there is strictly speaking no symmetry due to the random order of elements. The consequence of this is the main result of Fig.~\ref{fig:maps}, i.e. that the angular anisotropy of TDEs is heavily (although not completely) blurred by the random elemental ordering. In pure W, the low TDE events are found around the \hkl<100> and \hkl<111> directions (around 40 eV, in agreement with experiments, 42--44 eV~\cite{maury_frenkel_1978}) and the highest TDEs exceed 200 eV. Some directional dependence is still visible in the HEA, where on average directions close to \hkl<100> and \hkl<111> directions have lower TDEs, but the difference between low and high TDEs is only a few tens of eV (note the different scale of the colours in Fig.~\ref{fig:maps}). The maps of the standard deviation (Fig.~\ref{fig:maps} last column) show that in pure W, the TDE around e.g. the \hkl<100> direction is quite well-defined while for high-index directions the values can vary significantly due to thermal displacements affecting the probability for defect creation even for similar crystal directions. In contrast, the standard deviation map for the HEA is flat with values around 20 eV. This shows that the effect of random ordering of elements dominates over the randomness of thermal displacements. Interestingly, the standard deviations in the HEA (around 20 eV) is similar to the difference between average low-TDE events (30 eV) and average high-TDE events (50 eV) directions (Fig.~\ref{fig:maps} middle column.) Hence, the dependence on chemical environment is on average roughly as strong as the directional dependence.

\subsection{Mechanisms of defect creation}

\begin{figure}
    \centering
    \includegraphics[width=\linewidth]{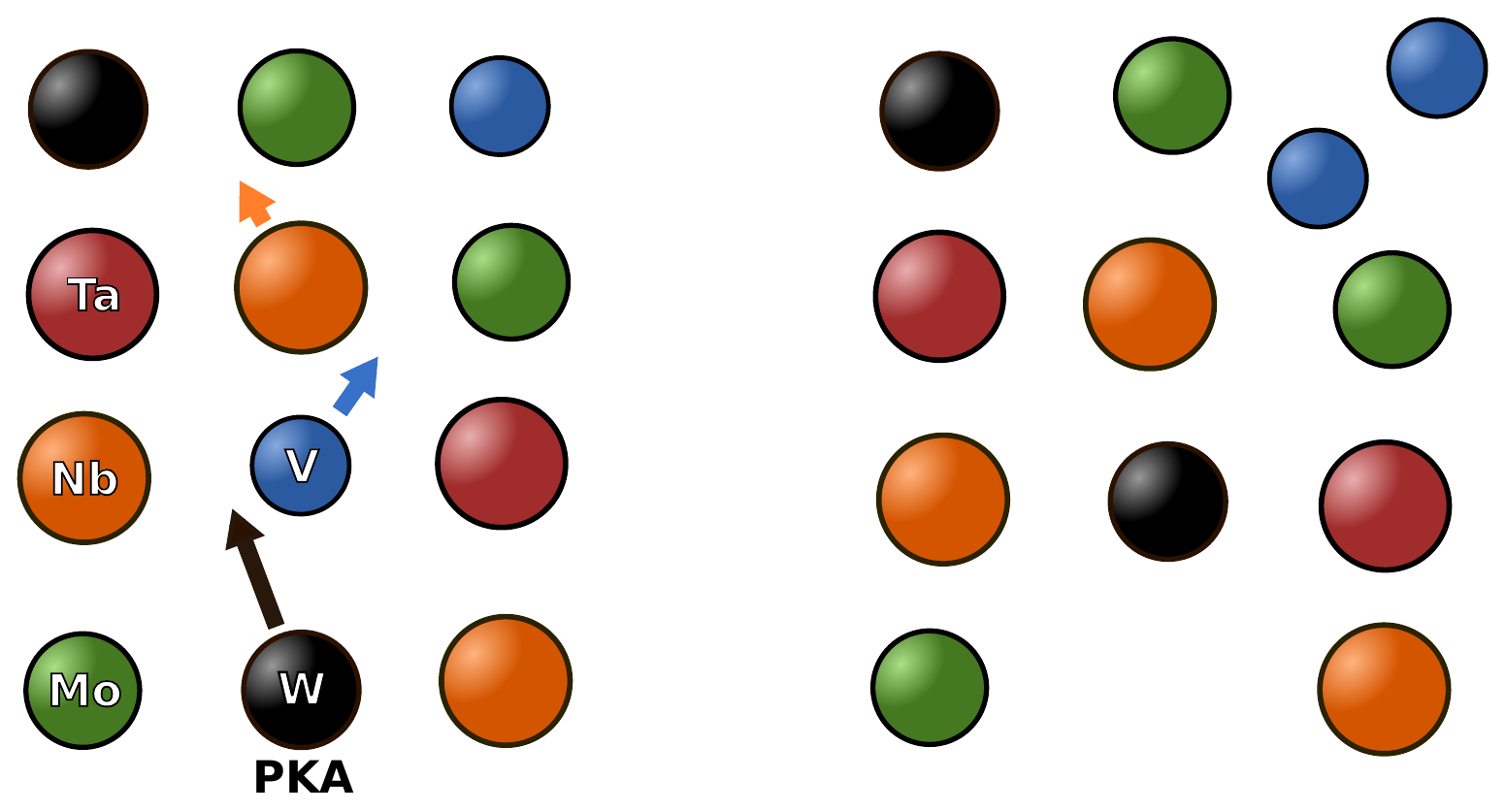}
    \caption{Schematic of a typical low-threshold event. A heavy W atom hits and displaces a light atom (V) and takes its lattice site. The displaced V atom is forced to find an interstitial site, which it typically does by forming a dumbbell with another nearby V atom.}
    \label{fig:mech}
\end{figure}

\begin{figure}
    \centering
    \includegraphics[width=\linewidth]{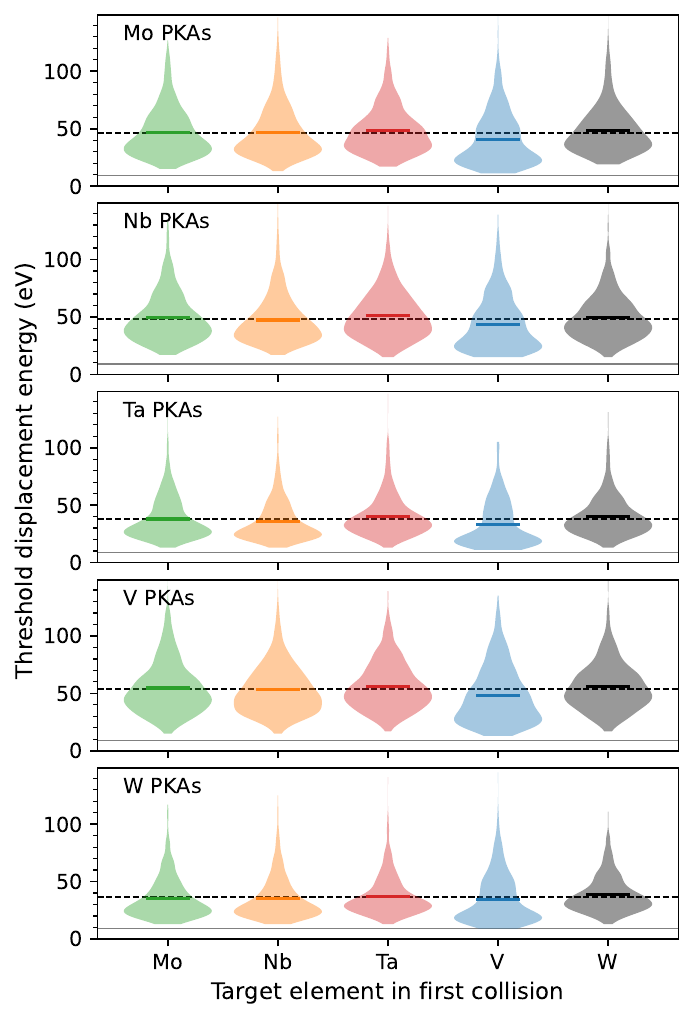}
    \caption{Violin plots showing distributions of the threshold displacement energies in the HEA separated by the element pair of the first major binary collision. The horizontal axis specifies the element of the first neighbour to be hit and receive a significant recoil from the primary knock-on atom. The dashed horizontal lines are the average of all simulations for each PKA element. The coloured lines inside the violins are the average values of the recoil pairs. The grey horizontal line indicates the global minimum threshold, 9 eV.}
    \label{fig:recoil}
\end{figure}

To better understand what controls whether a recoil along a given direction in the HEA has a high or low TDE, we studied the simulation trajectories of a set of low- and high-TDE events. As seen in Fig.~\ref{fig:hist}, low TDEs are as low as 10--20 eV, while high TDEs are well above 100 eV.

We find that low-TDE events are largely controlled by classical elastic collisions between different masses and conservation of momentum. Typical low-TDE events are when the PKA is an element with large mass (W) and it collides with a neighbouring small mass (V). Even with relatively low kinetic energy, the heavy element displaces the light mass and replaces it in its lattice site. The light element is then forced to find a stable interstitial configuration, which often is a V--V dumbbell. This is illustrated in Fig.~\ref{fig:mech} and in a few representative simulation animations in the Supplemental material. Additionally, in case of a both large and heavy atom (Ta), the probability of replacing the lattice site of a light element is further enhanced by the fact that large atoms in the HEA rarely even form stable interstitial configurations~\cite{byggmastar_modeling_2021}. The absolute lowest TDEs are found when this mechanism is combined with a low-TDE crystal direction such as \hkl<100> or \hkl<111>.

High-TDE events are more difficult to generalise. One could assume the same mechanism in reverse, i.e., that high TDEs are found when a light element collides but cannot displace a heavy atom and instead is recoiled back to its original lattice site. However, this only happens if the collision is nearly head-on so that the recoil is directed back. More often, the recoil trajectory after the first collision does not bring the PKA back to its lattice site and instead it may find and form a stable interstitial configuration elsewhere. The latter is especially likely if the PKA is V, as V--V dumbbells are by far the most stable interstitial configuration and very easily formed~\cite{byggmastar_modeling_2021,zhao_defect_2020}. Hence, light PKAs colliding with heavy neighbours may result in both low- and high-TDE events depending on the exact crystal direction and collision trajectory, and there is no clear general mechanism that describes a high-TDE event in the HEA. A few animations of high TDE events can be seen in the Supplemental material.

We analysed this more quantitatively by identifying the elements of the first major collision, i.e. the PKA and its approaching neighbour along the given recoil direction. Fig.~\ref{fig:recoil} shows distributions of the TDEs (as a 'violin plot') separated by the element pair of the first collision. Note that in many crystal directions, there is not a well-defined single neighbour atom that the PKA approaches. Here, we defined the first colliding neighbour as the first atom within 3.5 Å of the moving PKA that receives more than 10\% of the initial PKA energy. Tests proved this to be a reasonable criterion for identifying collision pairs. Even though in the case of many-body collisions obtaining only a pair of atoms is not fully representative, Fig.~\ref{fig:recoil} gives additional quantitative insight into the mechanisms of TDE events. In agreement with the explanation of low-TDE events above, it is clear that the TDE distributions for heavy elements (W, Ta) colliding with a light element (V) is strongly skewed towards lower TDEs compared to other distributions. High-TDE events are more difficult to identify and generalise, as argued above, although Fig.~\ref{fig:recoil} shows that the average TDEs when any element collides with one of the heaviest elements (W, Ta) is always higher than the global average (solid lines in Fig.~\ref{fig:recoil}).

\begin{figure}
    \centering
    \includegraphics[width=\linewidth]{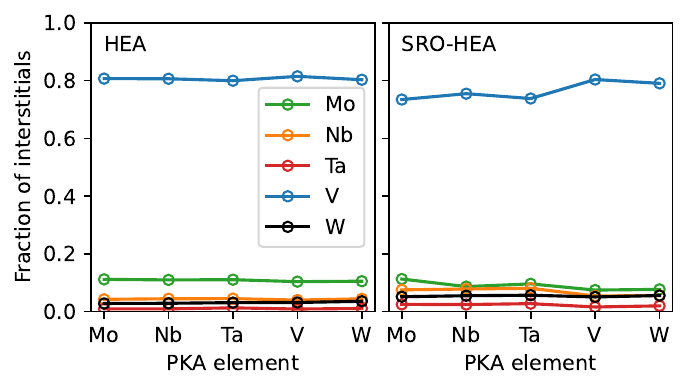}
    \caption{Elemental composition of interstitial atoms created in all TDE simulations in the HEA and SRO-HEA, as functions of the PKA element.}
    \label{fig:sia_comp}
\end{figure}

We also analysed the elemental composition of all interstitial configurations formed in the threshold events. Fig.~\ref{fig:sia_comp} shows the results, separated by the element of the PKA in both the random HEA and SRO-HEA. Based on the mechanisms explained above, in particular that heavy atoms displace light atoms, one could expect a corresponding bias in the composition of the final interstitial atoms (that heavy PKA atoms lead to light surviving interstitial atoms). However, Fig.~\ref{fig:sia_comp} shows that there is no dependence of the PKA element on the surviving interstitial elements in the HEA, and only small variations in the SRO-HEA. By looking at the simulation trajectories, this is explained by the same observation as in Ref.~\cite{byggmastar_modeling_2021}, that most stable interstitial configurations are V--V or other V-containing dumbbells. Even if other interstitial configurations are formed initially, they spontaneously and athermally relax to V-containing configurations since most other element configurations are not local minima in the potential energy landscape~\cite{byggmastar_modeling_2021}.

\subsection{Electron irradiation cross sections}

\begin{figure}
    \centering
    \includegraphics[width=\linewidth]{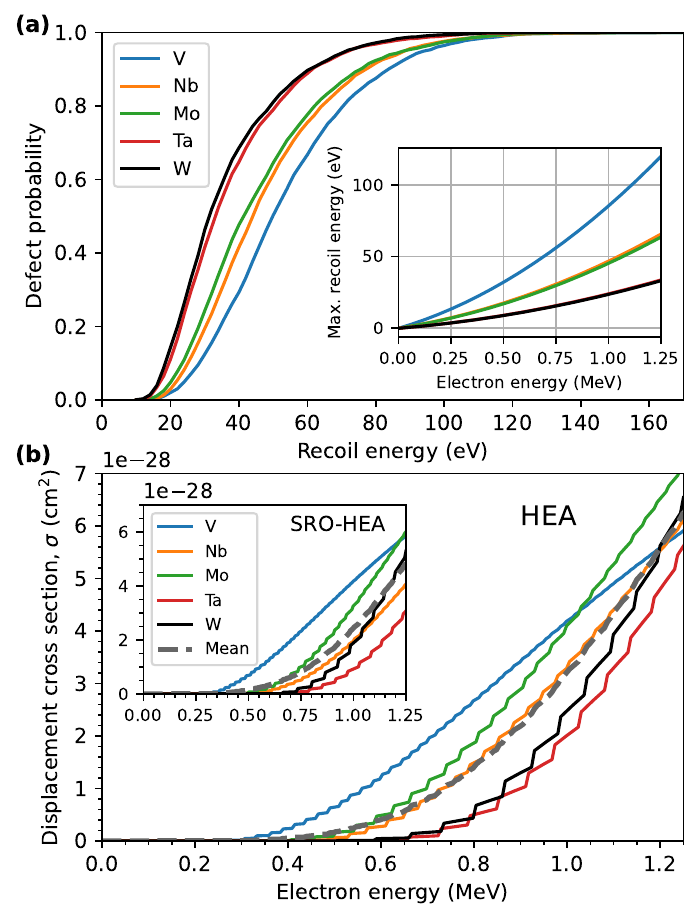}
    \caption{(a) Cumulative probability distributions for defect creation, separated by recoiling element. The inset shows the analytically calculated maximum energy transfer from electrons to each target element in the HEA. (b) Calculated displacement cross section as a function of electron energy in the HEA, and for the SRO-HEA in the inset. The dashed line is the arithmetic mean of all elements.}
    \label{fig:cross}
\end{figure}

Experimentally, the threshold displacement energies are obtained from the resistivity difference caused by Frenkel pairs created in electron irradiation. We have found that on average, recoils of the lightest element (V) have the highest TDE ($53.5 \pm 0.3$ eV in the random HEA) and the heaviest element (W) the lowest TDE ($36.2 \pm 0.3$ eV). However, in electron irradiation experiments, the maximum kinetic energy transfer would be highest for collisions with light target elements. Hence, V would receive higher recoil energies than W, which may disguise the fact that the TDEs for V recoils is also higher. To investigate this, we calculate the cross section for defect production by numerically integrating (following Refs.~\cite{lucasson_production_1962,nordlund_molecular_2006}):
\begin{equation}
    \sigma (E) = \int_0^{T_\mathrm{max}} P_\mathrm{d} (T) \ \mathrm{d} \sigma (T).
    \label{eq:cross}
\end{equation}
Here, $\sigma (E)$ is the cross section for the electron energy $E$, $T$ is the recoil energy of the PKA, $T_\mathrm{max}$ is the maximum energy transfer from the electron (with energy $E$) to the PKA. The differential cross section for giving an atom the recoil energy $T$, $\mathrm{d}\sigma (T)$, can be estimated with the McKinley-Feshbach approximation~\cite{mckinley_coulomb_1948}, eq. (20) in Ref.~\cite{lucasson_production_1962}. $T_\mathrm{max}$ is calculated with the relativistic equation for maximum energy transfer in a binary collision
\begin{equation}
    T_\mathrm{max} (E) = \frac{2ME(E+2mc^2)}{(m+M)^{2}c^{2}+2ME},
\end{equation}
where $c$ is the speed of light and $m$ and $M$ are the masses of the electron and the lattice atom, respectively. The inset in Fig.~\ref{fig:cross}a shows $T_\mathrm{max} (E)$ for the different elements in the HEA.

$P_\mathrm{d} (T)$ in Eq.~\ref{eq:cross} is the defect probability distribution, which we obtain directly from the TDE simulations and is shown for each recoiling element in Fig.~\ref{fig:cross}a. Strictly speaking $P_\mathrm{d} (T)$ should not be the probability of forming a defect with recoil energy $T$, but the average number of defects formed, $N_\mathrm{d} (T)$~\cite{nordlund_molecular_2006}. At higher $T$, more than one Frenkel pair can form, and all of them would contribute to the resistivity measured experimentally. However, since in our simulations we stopped the recoil energy increment when the TDE was found for a given direction, we do not have access to the true $N_\mathrm{d} (T)$, but instead only know whether any defects are formed or not, i.e. $P_\mathrm{d} (T)$. Fortunately, the cross-section integral value is dominated by low $T$, where $P_\mathrm{d} (T)$ should not differ much from $N_\mathrm{d} (T)$. 

Fig.~\ref{fig:cross}b shows that indeed even though V recoils need the highest threshold energies, in electron irradiation they will be the first and most likely recoils to produce defects when increasing the electron beam energy. The order of displacement cross sections is thus completely reversed compared to the order of raw threshold displacement energies. (compare Figs.~\ref{fig:cross}a and b). In reality, the electrons collide with all elements with equal probability in the equiatomic HEA, creating defects according to an average of the element-specific displacement cross sections. Fig.~\ref{fig:cross} also shows this average cross section as a dashed line. We can conclude that if an electron irradiation and resistivity measurement for the HEA were to be carried out, stable Frenkel pairs would start forming mainly by V recoils at around 0.3--0.4 MeV electron energies.

\section{Summary and conclusions}

We have comprehensively simulated and analysed the statistics and mechanisms of threshold displacement events in equiatomic MoNbTaVW high-entropy alloys using a machine-learned interatomic potential. The average threshold displacement energy is $44.3 \pm 0.15$ eV, which is significantly lower than in any of the constituent pure metals. The effect of short-range ordering is fairly weak, causing a modest increase of the average TDE to $48.6 \pm 0.15$ eV. We observed that the main mechanism for defect creation at low recoil energies is defined by simple conservation of momentum when a heavy atom collides with a light atom, displacing and replacing it. The lowest threshold displacement energies are hence obtained when the primary recoil is a heavy atom (W, Ta). However, when considering the cross sections for defect production in electron irradiation, due to the mass-dependent energy transfer V will still be the first recoils to generate defects when increasing the electron beam energy. 

We conclude that while many material properties of refractory high-entropy alloys follow Vegard's law, i.e. the composition-weighted average of the pure metals, this is far from true for threshold displacement energies which depend on the local chemical environment. The mechanisms and trends observed here for MoNbTaVW are likely also true for other refractory alloys that combine elements of very different size and mass. Our results combined with simple damage models suggest that the extent of the primary damage is higher in refractory high-entropy alloys compared to pure refractory metals.

\section*{Acknowledgements}

The work was supported by funding from the Research council of Finland through the OCRAMLIP and HEADFORE projects, grant numbers 354234 and 333225. This work has been carried out within the framework of
the EUROfusion Consortium, funded by the European Union via the Euratom Research
and Training Programme (Grant Agreement No 101052200 — EUROfusion). Views and
opinions expressed are however those of the author(s) only and do not necessarily reflect
those of the European Union or the European Commission. Neither the European Union
nor the European Commission can be held responsible for them. Grants of computer capacity from CSC - IT Center for Science are gratefully acknowledged. The authors wish to thank the Finnish Computing Competence Infrastructure (FCCI) for supporting this project with computational and data storage resources.

\bibliography{mybib}

\end{document}